# Space-time Bayesian analysis of the environmental impact of a dismissing nuclear power plant

A. Petraglia[1], C. Sirignano[1], R. Buompane[1], A. D'Onofrio[1], A. M. Esposito[2], F. Terrasi[1], C. Sabbarese[1]

[1]CIRCE, Dipartimento di Matematica e Fisica, Università degli Studi della Campania "L. Vanvitelli" Caserta, Italy
[2]Sogin, Garigliano NPP, Sessa Aurunca (Caserta)

*Abstract*

The assessment of the radiological impact of decommissioning activities at a nuclear power plant requires a detailed analysis of the distribution of radionuclides in the environment surrounding it. The present work concerns data of three campaigns carried out during the last twenty years in the plain of the Garigliano river surrounding the Garigliano Nuclear Power Plant (GNPP), which is located in Southern Italy and shut down in 1979. Moreover, some data from surveys held in the eighties, across the Chernobyl accident, have been taken in account. The results for the soil samples, in particular for $^{137}$Cs and $^{236}$U specific activity, were analyzed for their extension in space and in time. Some of the problems related to the classical analysis of environmental radiological data (non-normal distribution of the values, small number of sample points, multiple comparison and presence of values lesser than the minimum detectable activity) have been overcome with the use of Bayesian methods.

The scope of the paper is threefold: (1) to introduce the data of the last campaign held in the Garigliano plain; (2) to insert these data in a larger spatio-temporal frame; (3) to show how the Bayesian approach can be applied to radiological environmental surveys, stressing out its advantages over other approaches, using the data of the campaigns.

The results show that radionuclides specific activity in soil is dominated by the natural sources with the contribution of the atmospheric fallout. A detailed study was performed on the $^{137}$Cs data to evaluate both their statistical distribution and the trend over the space and the time. It results that (i) no new contribution there was in the last decades, (ii) specific activity values of the area surrounding the GNPP are consistent with those obtained in other farther areas, (iii) the effective depletion half-life factor for $^{137}$Cs is much lower than the half-life of the radionuclide.

## 1. Introduction

The decommissioning of a Nuclear Power Plant (NPP), at the end of its productive cycle, often presents more societal issues than radiation protection problems. The needed knowledge and techniques to accomplish the series of operations of decontamination of the technological structures and buildings of the plant and their removal under safety conditions often are fully acquainted, but the aspects related to the impact on the public opinion still need full attention. Since 2000 a group of physicists of the University of Campania "L. Vanvitelli" (before Second University of Naples) has started a research program in collaboration with the engineers of the Sogin (Society for the management of the Italian nuclear plants) working on the Garigliano Nuclear Power Plant (GNPP), located in Southern

Italy and shut down in 1979. The aims of this collaboration were verifying and updating the survey methodology, assessing the impact of decommissioning on the environment and communicating the results to the society, in order to improve the awareness on the activities carried out on the site. In particular, the state of the environment radioactivity in the area surrounding the GNNP has been recorded over the years with different radio-ecological campaigns, aiming to acquire the current situation and to address the sources of both natural and anthropic radioactivity.

Statistical analysis plays a fundamental role for the investigation of data related to pollutants in the environment. Among other analyses the comparison between different groups is of particular importance. The groups may refer to different geographical areas, periods of measurement, population groups, and so forth. However, the values corresponding to a given group show a statistical variation from sample to sample, particularly for small groups, which open a number of possibilities to the experimenter regarding the comparison with other data sets. The commonly used methods are based on qualitative analyses and significance test of the null hypothesis: t-test and analysis of variance, belonging to the frequentist approach.

In this work, a reconstruction of the spatial and temporal variations over more than 30 years, after the NPP stop, is presented using the Bayesian approach. We found the Bayesian approach more straightforward and simpler than the frequentist approach in solving the issues related with the application of frequentist methods to non-normal data distributions, small groups, and the presence of values below the minimum detectable activity [1,2].

Spatial variations are studied by collecting samples in several points of the plain of the Garigliano and in some points of the Sele plain, geologically similar to the Garigliano plain, but geographically far away (about 130 km). The temporal analysis was performed based on the results of five campaigns carried out in the surroundings of the Garigliano NPP beginning from 1985.

## 2. Materials and Methods

**Sampling campaigns and measurement techniques**

The GNPP is located in central Italy, about halfway between Rome and Naples. It definitively stopped its activities in 1986.

Since the year 2000, three campaigns were carried out in the surroundings of the GNPP by the University of Campania "L. Vanvitelli" in collaboration with Sogin: in 2001/2002 [3] in 2008/2009 [4] and in 2015/2016 [5], here named *camp2002*, *camp2008-12* and *camp2015*, respectively. Every campaign was planned independently from the precedent results. As comparison, data from two previous environmental campaigns held by ENEL [6] were also included. These early campaigns, conducted in 1985 (*camp1985*) and in 1986 (*camp1986*) embrace the Chernobyl accident, occurred in April 1986. The samples of these campaigns have been collected on the whole plain.

The parameter chosen for the comparison analysis presented in this study is the $^{137}$Cs specific activity measured on superficial (i.e. belonging to the first 15-20 cm) unspoiled soils samples. A total of 234 samples were selected among all other kind of samples and measurements collected over all the campaigns. Uncultivated soil samples, collected in the first 15 cm and evenly distributed in the study area, were selected among those investigated during the first two campaigns. Table 1 shows a summary of the informations about the surveys.

| Location | Campaign | Area | n. of samples | Ref. |
|---|---|---|---|---|
| Garigliano | camp1985 | | 8 | [6] |
| Garigliano | camp1986 | | 8 | [6] |
| Garigliano | camp2002 | | 48 | [3] |
| Garigliano | camp2008-12 | A | 17 | [4] |
| Garigliano | camp2008-12 | B | 8 | [4] |
| Garigliano | camp2008-12 | C | 7 | [4] |
| Garigliano | camp2008-12 | D | 11 | [4] |
| Garigliano | camp2008-12 | E | 8 | [4] |
| Garigliano | camp2015 | A | 6 | [5] |
| Garigliano | camp2015 | B | 24 | [5] |
| Garigliano | camp2015 | C | 13 | [5] |
| Garigliano | camp2015 | D | 22 | [5] |
| Garigliano | camp2015 | E | 34 | [5] |
| Sele | camp2008-12 | F | 7 | [4] |
| Sele | camp2015 | F | 13 | [5] |

*Table 1. Summary of the campaigns subject of the study.*

A large number of samples was collected during the two most recent campaigns. Their distribution around the site of interest allowed a further distinction in 6 different areas: area *A* is centered on the GNPP and it has a 2.2 km radius; other areas cover a 90° sector each, outside area *A*, and are labeled alphabetically from *B* to *E*, clockwise from the south-western sector (Figure 1). The letter *F* refers to the control area located roughly 130 km to the SE of the NNP: the Sele Plain. No distinction in areas have been done for the 1985, 1986, and 2002 campaigns.

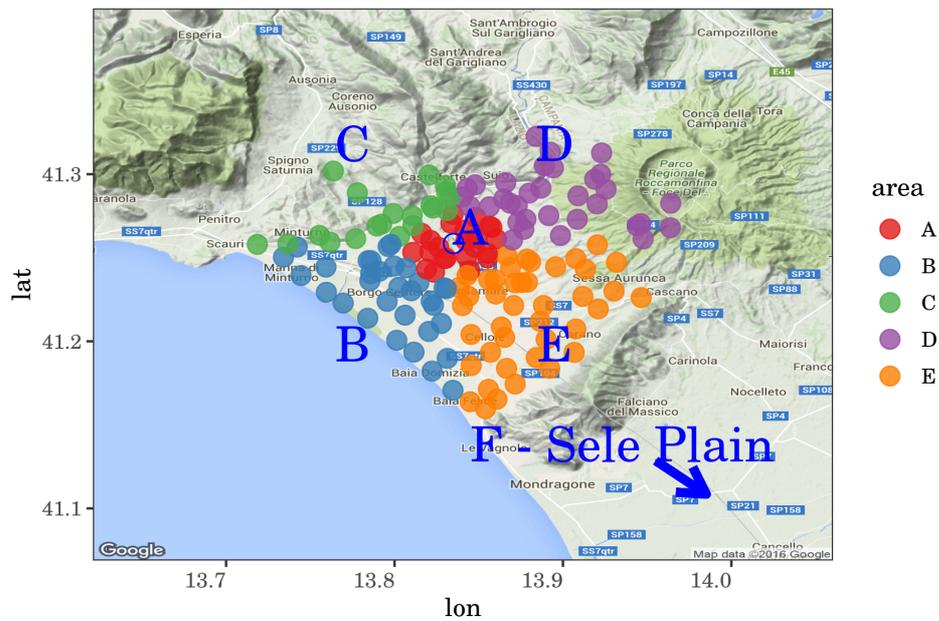

*Figure 1: Map showing the surroundings of the Garigliano NPP (blue circle). Each full-colored circle identifies the position of a superficial soil sample collected during camp2008-12 and camp2015. A different color code is assigned according to the area of belonging. The control area (Sele Plain) has been labeled with the letter F and it is not shown in the map.*

The specific activities of $^{137}$Cs on the samples were measured using high energy resolution and low background γ spectrometry.

A high-resolution germanium hyperpure γ-ray detector (1.9 keV resolution at 1.332 MeV and 70% efficiency) properly shielded was used for the measurement of the samples belonging to camp2002, camp2008-12 and camp2015. The spectra were acquired, displayed and analyzed on computers running Silena Gamma+ and Ortec GammaVision. Typical sample volume was 8000 cm$^3$ (20 cm x 20 cm x 20 cm) of soil. Samples were dehydrated, homogenized, sieved and put in Marinelli vessels for analysis. The efficiency calibration curves and minimum detection activities (MDA) were accounted for.

Soil samples collected during *camp2015* were also analyzed for the determination of $^{236}$U abundances by Accelerator Mass Spectrometry (AMS) facility at Circe, consisting of a NEC 9SDH-2 Pelletron accelerator operated at 3 MV for $^{238}$U acceleration and scaled for the same magnetic rigidity for the other U isotopes. The addition of a known amount of an isotope of the same element missing in the sample (spike) enables the determination of the concentration of ultratrace contaminants, by comparing the relative abundances. The rare isotopes ($^{236}$U and the $^{233}$U spike) are counted in an ionization chamber. The methodology to isolate and concentrate U was adapted from method No.ACS07 (Eichrom Technologies) reported by Eichrom analytical procedure. Each solution containing typically 10 g of dissolved bulk sample was spiked with a $^{233}$U standard (IRMM058). U separation was performed by Eichrom UTEVA resin. Columns of 5 ml of UTEVA resin (1 g) were used. The eluate was dried and U precipitated as aluminum salt by adding Al(NO$_3$)$_3$ in acid solution taken to dryness. After combustion in a muffle furnace at 800 °C for 8 h, samples were pressed in aluminum cathodes.

Typically, the samples here analyzed were measured as "U-poor" samples according to the method described in [7]. A carousel holding 40 samples (including standard and blank cathodes) was mounted in the cesium sputter ion source. Several cathodes containing known amounts of two standards of Uranylnitrate with a certified $^{236}$U/$^{238}$U ratio of (6.98 ±

0.32)10⁻¹¹ (Vienna KkU]) and (1.06519 ± 0.00075) 10⁻⁸ (IRMM-075/5) were used for normalization and quality check.

**Data Analysis**

In our approach, we addressed issues related with the application of classical (frequentist) statistical analysis to different data distributions and in the presence of values below the minimum detectable activity (MDA).

For this work the comparison groups were the samples taken in areas A through F (for the last two campaigns) and in different campaigns held in the last decades. There were, for some campaigns and areas, just a few points. This is, in general, a problem for frequentist tests, that are known to be biased for small groups. However, this is not a big problem for Bayesian analysis, that, with right specifications, can outperform frequentist analysis, for small groups: indeed, this is one of the advantages of the Bayesian approach over the classical analysis. The calculations, tests, and the diagrams were performed using the software R [8] in the framework Rstudio [9].

In order to take any test between sets of data, it is necessary to define the statistical distribution of values in each group. In fact, the tests to be performed are very sensitive to the distribution of data [10].

The check of the distribution was performed by means of various statistical tests (Shapiro, Kolmogorov-Smirnov and others), which test the null hypothesis against the assumption of normality. However, care should be given, otherwise the test can become unreliable. Indeed, example can be given, where non-normal small data sets are detected as normal, or normal distributed big data sets are detected as non-normal.

In cases in which the distribution is absolutely unknown, we can use the so-called non-parametric tests, which do not make any assumption on the distribution. They provide a greater strength in presence of of outliers, but at the expense of a lower specificity (i.e. the ability to discern differences).

However, for contaminants in the environment, we can find a distribution that approximates the data acceptably. Various studies have examined this distribution, some of them [11–13] have shown, both theoretically and experimentally, that in the presence of dilution, a soluble contaminant tends to create concentrations that are log-normal. This distribution is such that its logarithm is normally distributed: $ln(x) \sim N(\mu, \sigma)$, where $N$ is the Normal distribution.

Blackwood [11] analyzes the implications of using the log-normal distribution to radiological monitoring applications. He also considers the robustness of the log-normal when used for other distributions, such that gamma, mixtures of log-normal, normal. He concludes that such a choice has little penalty in terms of accuracy.

The parameters of a log-normal distribution are related to the mean and variance of the data as:

$$\mu = \log(mean) - \frac{1}{2}\log\left(1 + \frac{variance}{mean^2}\right) \quad (1)$$

$$\sigma = \sqrt{\log\left(1 + \frac{variance}{mean^2}\right)} \quad (2)$$

Note that $e^\mu$ corresponds to the median of the data. The mean is:

$$mean = e^{\left(\mu + \frac{\sigma^2}{2}\right)} \quad (3)$$

and

$$variance = e^{(2\mu + \sigma^2)}\left(e^{\sigma^2} - 1\right) \quad (4)$$

We will show below that the specific activity of $^{137}$Cs in the environment is modeled very well by a log-normal distribution.

**Multiple comparisons**

The classical method of comparison between two groups with normal distribution is the two-sample t-test. If the distribution of two sets of data is normal, the difference between them will have a t-student distribution and the null hypothesis corresponds to assume the difference between the averages equal to zero. For more than two groups the analysis of variance (anova) can be used.

The result obtained, however, is not credible because the basic hypothesis, that is the normality of the two distributions, is not verified. The investigator has, thus, two choices:
- the use of non-parametric tests;
- the use of a distribution that better approximates the data.

However, both from theoretical considerations and data analysis, we suppose that the activity data have a log-normal distribution. From the practical point of view the comparison procedure consists in performing a t-test (or anova) on the transformed values: *x' = ln(x)*.

There is also another problem: for multiple comparisons, such as comparisons for many control factors, the probability of erroneous inferences increases [14].

We solve all of these problems with the Bayesian approach. Indeed, both the t-test and analysis of variance belong to the frequentist approach. They significantly depend on the p-value, which is the probability of the given or a more extreme outcome, if the null hypothesis is true. If this probability is small, then the null hypothesis is rejected, otherwise we fail to reject it. This procedure is somewhat convoluted and often leads to its misuse [15]. In this work we do not use this approach.

In the Bayesian approach the parameters of a set of data (mean, standard deviation) are considered random variables. Thus, they have a distribution which can be analyzed. For example, in comparisons we consider the differences of the posterior distributions.

Thus, starting from the inferred distribution (the so-called *prior*: $p(\theta)$, where θ is the set of distribution parameters) and from the data (the *likelihood*: $p\langle D|\theta\rangle$, which is the probability of obtaining the data *D*), the expected distribution, the so called *posterior* is calculated.

$$p\langle\theta|D\rangle = \frac{p\langle D|\theta\rangle\, p(\theta)}{\int p\langle D|\theta\rangle\, p(\theta)\, d\theta}$$

The integral in the denominator, defined on the whole range of θ, is a normalizing term for $p\langle\theta|D\rangle$.

The calculated posterior, is a set of values having the same distribution of the population data, on which we can make our analysis (basic inference, model checking, prediction and so on). This approach greatly simplifies some conceptual and practical issues. For example, the null test on the parameter θ is reduced to verify that its confidence interval includes the zero [2].

This simplification is obtained at the expense of a numeric complication. This, however, is no longer a problem, thanks to the computing power of modern computers and some theoretical developments. For complex analyses (many groups, the presence of group nesting) the Bayesian approach can give many advantages, some of them directly concerning the environmental radionuclide monitoring problems.

The model is thus

$$A_j = LogNorm(\mu_{ik}, \sigma_{ik})$$

In which we choose:

- log-normal likelihood (*LogNorm*):
  $A_j$ (j goes from 1 to number of samples) log-normal distributed with mean $\mu_i$ and standard deviation $\sigma_i$.
- Non informative priors, i.e. with very wide distribution:
  $\sigma_{ik}$ (standard deviation) evenly distributed (flat prior).
  $\mu_{ik}$ (mean) normally distributed (gaussian prior).
  The indexes $i$ and $k$ in $\mu_{ik}$ and $\sigma_{ik}$: discriminate the different groups, in our case the areas or the campaigns, respectively. This provides a different parameter set for each group.

Often, in the analysis of environmental radioactivity, we have to deal with values smaller than the MDA of the measurement procedure, also called censored or nondedect data. This is a general problem that appears in the study of environmental pollutants. Some classical approaches for the inference of the missing values are reported in the literature [16,17]. They are, however, little used because quite complex. In fact it is very common to treat the nondetect data drastically: they are not considered in the calculation of statistics (means, variances, etc.) or they are all approximated to MDA (as a conservative approximation) or half MDA. These solutions are, however, generally unsatisfactory [18].

The Bayesian approach for the comparison of different groups, once implemented via software, can straightforwardly be extended so as to account for values lower than MDA, by inferring them according to a predetermined distribution and including them in the comparison [2]. To this scope, in the following, brms, an R package for he Bayesian modeling, including modeling of log-normal distributions, censored data, group-level parameters, and so on, has been used [19].

This approach has been used in some cases of chemical pollution (such as pesticide pollution [20,21]) and a few times applied to radioactivity measurements [22,23].

In our case we got just one value less than MDA, however, in a work of one of the authors [24], the adequacy of the method has been demonstrated in a case study with high proportion of MDA values.

## 3. Results

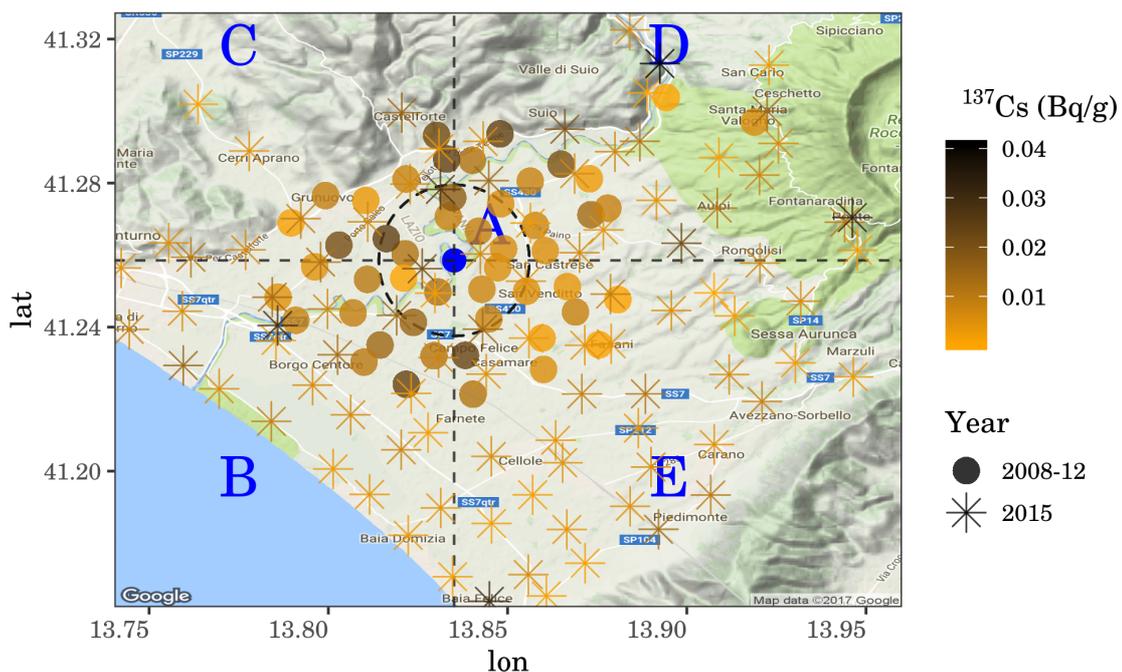

*Figure 2: Specific activity, in false colors, of $^{137}Cs$ in soils collected in the last two campaigns.*

## The $^{236}$U and $^{137}$Cs specific activities

The specific activity of $^{137}$Cs in soils samples collected in the last campaigns is shown in Figure 2. No clear spatial differences seem to exist for different areas, suggesting a global origin, such as fall-out of nuclear tests or Chernobyl accident [25,26]. However, a deeper, more formal analysis is necessary, as will be shown below.

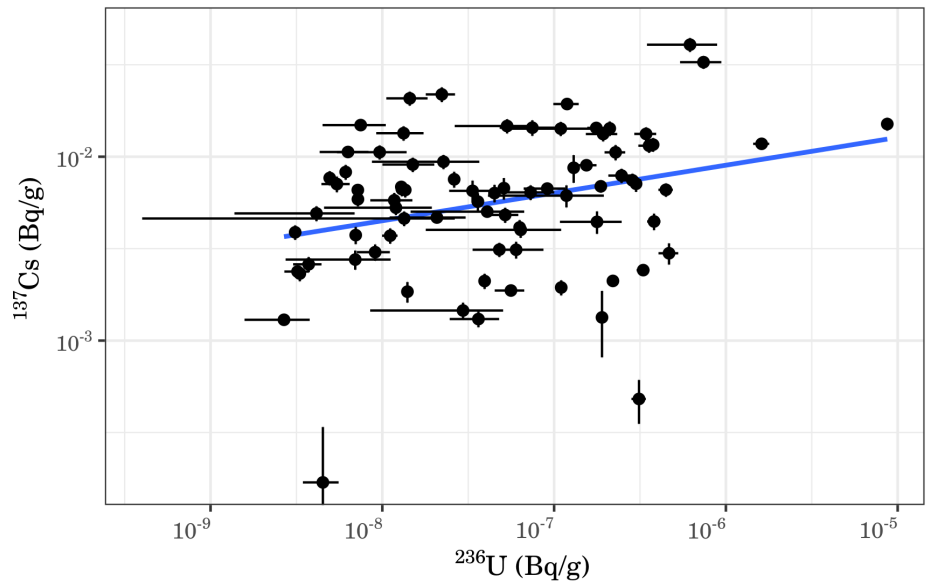

*Figure 3: Comparison of the specific activity of $^{137}$Cs and $^{236}$U (log-log scales).*

Various studies have been made in the last years about the presence of $^{236}$U in the environment. For a recent review, see [27]. In the environment, at least four mechanisms/source can be proposed, some of them are anthropogenic, due to contamination from nearby nuclear facilities and global fallout [28–33]. The comparison of the $^{236}$U and $^{137}$Cs is shown in Figure 3 (log-log scale). The values vary across some order of magnitude in both axis and show a wide oscillation, due to intrinsic properties of log-normal distributions. The slope is 0.14 ± 0.08. Thus, the 95% interval includes zero, therefore, we should reject the hypothesis of a positive correlation between the two radionuclides. However, further studies and sampling are still ongoing to exclude type II error.

**The statistical distribution of specific activity in the environment**

A fundamental role in the analysis of the data is played by the modeling of the distribution. Studies have shown that in the presence of dilution, a soluble contaminant in the environment tends to create concentrations that are log-normal.

This is true for our $^{137}$Cs data. Indeed, in Figure 4, the experimental specific activity is compared to Normal and log-normal distributions, with mean and standard deviation calculated from the data, for each different campaign. The latter shows a better fit.

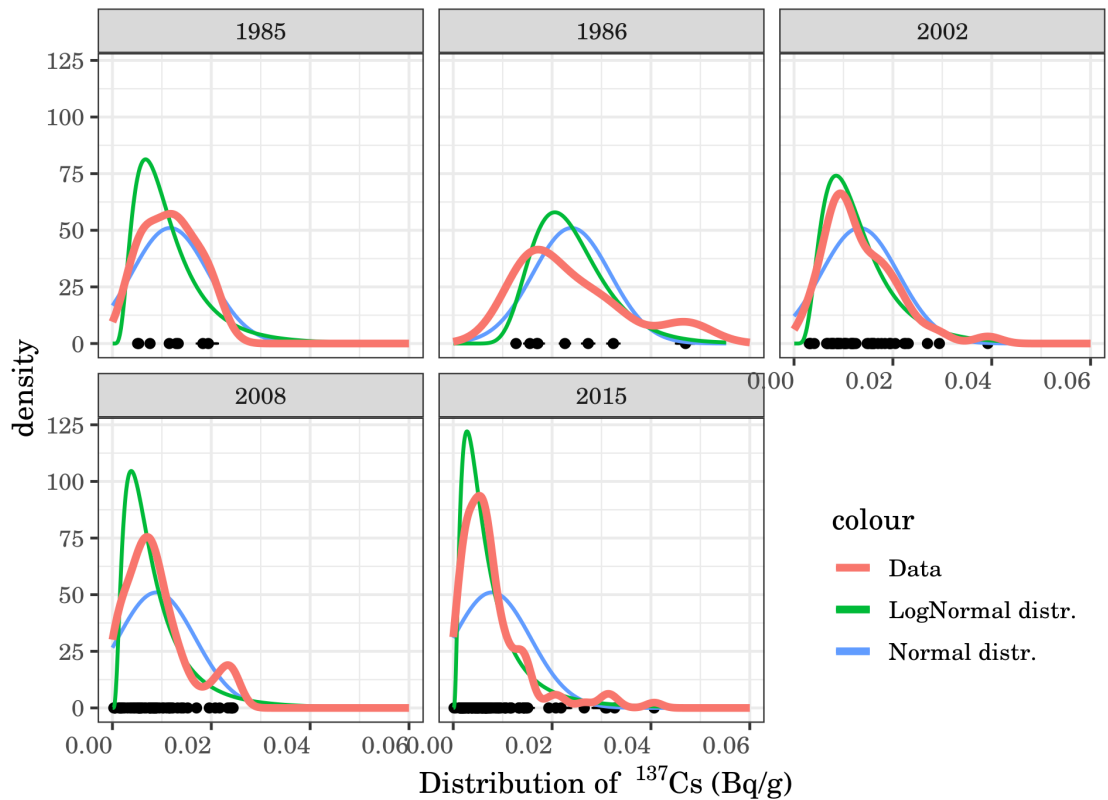

*Figure 4: Comparison of the experimental data and Normal and log-normal distributions, with mean and standard deviation calculated from the data, for the different campaigns.*

**Comparison between different areas and campaigns**

For the calculations we used the *rjags* and *brms* libraries in R. They use Markov Chain Monte Carlo (MCMC) methods to draw samples from the posterior distributions [2,19,34]. The values (just one in this case!) smaller than the Minimum Detectable Activity (MDA) have been handled following the procedure explained above; this number will likely increase in the next years, due to the depletion of the specific activities and the hopefully absence of new releases (see "behavior in time" below).

Figure 5 shows a set of calculated posterior distributions of the $^{137}$Cs specific activity, for all data merged together, compared with the actual data distribution, in blue. Both the normal and log-normal fit are shown for comparison.

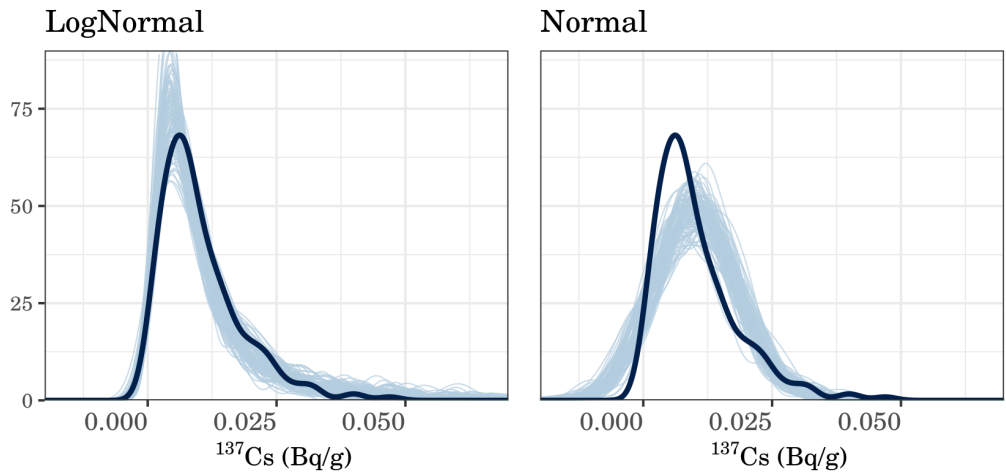

*Figure 5: Calculated posterior distributions of the $^{137}$Cs specific activity for all data merged, compared with the actual data distribution, in blue. Both the normal and log-normal fit are shown.*

The posterior distribution for each area has also been calculated. Their comparison is shown in Figure 6. There is always a superposition between areas, demonstrating no evident differences of the specific activity in the environment. This can be better seen by direct comparison of the 95% HDI (Highest Density Interval [2]) of all the zones, which is shown in table 2.

The lack of differences between areas is a clear hint of no correlation of the specific activity with distance from the NPP.

| *Campaign* | *Area* | *hdi_min (Bq/g)* | *hdi_max (Bq/g)* |
|---|---|---|---|
| 2008 | A | 0,0048 | 0,0100 |
| 2008 | B | 0,0047 | 0,0142 |
| 2008 | C | 0,0042 | 0,0136 |
| 2008 | D | 0,0040 | 0,0101 |
| 2008 | E | 0,0018 | 0,0055 |
| 2008 | F | 0,0029 | 0,0067 |
| 2015 | A | 0,0026 | 0,0095 |
| 2015 | B | 0,0025 | 0,0143 |
| 2015 | C | 0,0024 | 0,0149 |
| 2015 | D | 0,0029 | 0,0169 |
| 2015 | E | 0,0020 | 0,0104 |
| 2015 | F | 0,0013 | 0,0106 |

*Table 2. HDI 95% limits of the $^{137}$Cs distribution for each area and for the last two campaigns, showing superposition between all of them.*

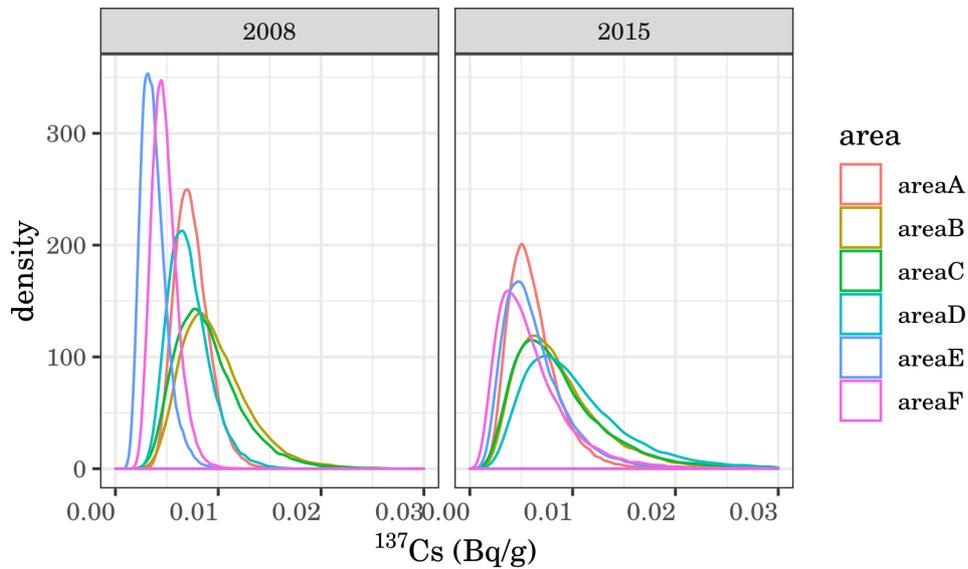

*Figure 6: Comparison of the posterior distribution for each area.*

Moreover, combinations of the parameters, from the sample posterior distributions, can be created right away. In fact, as mentioned above, in the Bayesian approach they are random variables, and thus themselves have a distribution. This allows to overcome the null hypothesis test by direct analysis of the distribution difference, as is demonstrated in Figure 7, where the difference between the two most different areas of the 2015 campaign, D and F, is shown.
In it, the 95% HDI (the blue thick line at the base of the graph) includes zero, thus indicating that we can accept the hypothesis of equality of the values of activity of the zones D and F.

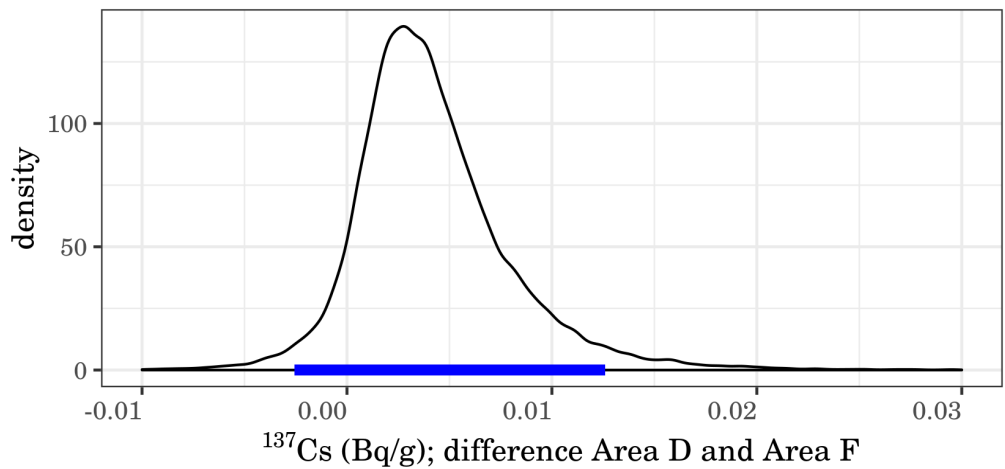

*Figure 7: Distribution of difference between the two most different sampling areas, D and F, is shown. The 95% HDI is also shown as blue thick line at the base of the graph.*

**Behavior in time**

The time extent of the data covers a range of thirty years. This allows to make a time analysis of all the data. The scope is to address possible anomalous contaminations and to determine the parameters of the depletion curve.

Figure 8 shows the time behavior of the $^{137}$Cs specific activity (in log y-axis) Vs. time (year). The brown boxes/whiskers show the experimental data. The lower and upper hinges correspond to the first and third quartiles. The whiskers extend from the hinge to the values no further than 1.5 * IQR from the hinge, where IQR is the distance between the first and third quartiles. The only outlier is plotted individually. Blue dots indicate the mean of the posteriors calculated values. A regression curve through them (from 1986 on) is also shown in blue. A discontinuity of the specific activity is evident between 1985 and 1986: it can be likely attributed to the Chernobyl accident, happened in the spring 1985, and its subsequent fallout [25,26].

No clear discontinuity turns out to exist between the campaigns held in 2008 and 2015, across the Fukushima accident (2011). This proves no clear effect, as expected, due the considerable distance between the Garigliano and Fukushima NPPs.

The green line in figure, is the expected depletion of the 1986 $^{137}$Cs, supposing only a radioactive decay, with half life of 30.17 years. The difference is caused by the geochemical processes of the Cs in soils. The effective half-life, calculated from the data, is 13.89 years.

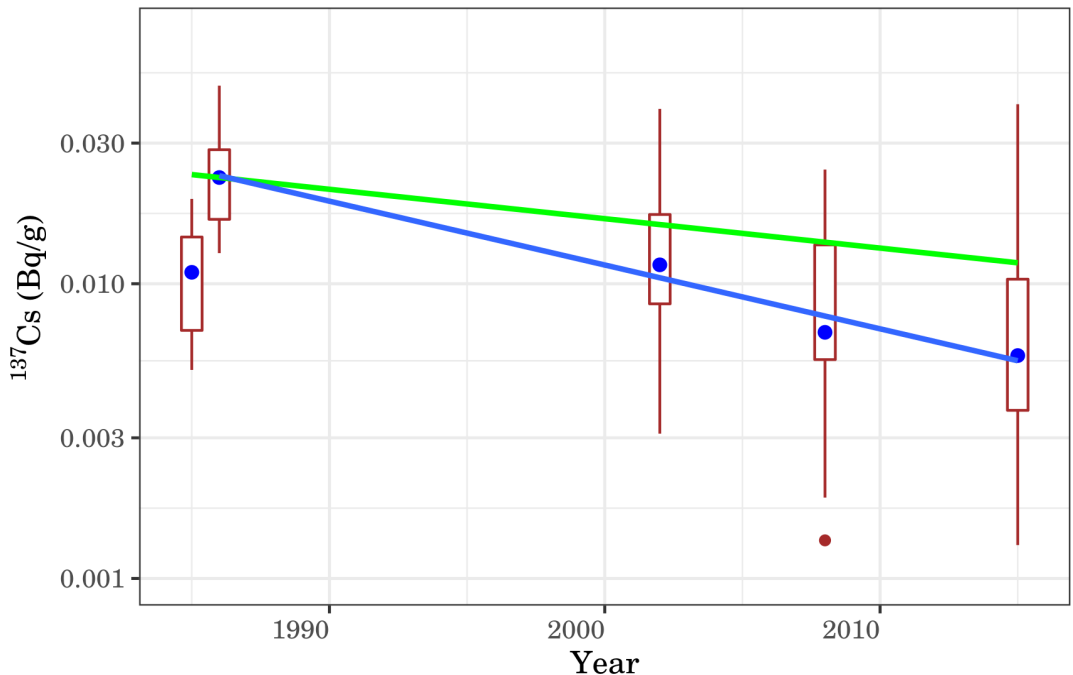

*Figure 8: Behavior of the $^{137}$Cs specific activity (in log y-axis) Vs. year. The discontinuity in 1985 is clearly visible, likely due to the Chernobyl accident and its subsequent fallout. The green line is the decay behavior, if due only to radioactive decay. The real line, in blue, shows a steeper descent, due to concurrent geochemical processes.*

## 4. Discussion and conclusions

The assessment of the environmental impact of the dismissing GNPP has been made by measurements of the specific activity of anthropogenic radionuclides in soils. The scope was to check the environmental status of the area surrounding the NPP, the origin of anthropogenic radioactive isotopes and the effects on the neighboring population health.

The possible presence of former radioactivity of anthropogenic origin has been addressed by measurements in surrounding and control areas and along more than three decades in time.

In this work we show the results involving superficial soils. They show always an overall low specific activity with no area dependencies, including the control area.

The anthropogenic activity was compatible with the fall-out of cold war nuclear tests and of the Chernobyl accidents. No measurable effects from the Fukushima accident or other GNPP activities could be measured in the areas under study.

Using the Bayesian approach in the radiological study of environmental samples, have demonstrated some clear-cut benefits:
- the simple conceptual framework;
- the good behavior for small samples;
- the simplicity of the comparisons across groups;
- the possibility to include different prior distributions;
- the possibility to deal with nondetects values.

This approach confirms and strengthen the picture above and some previous work, and gives more precise numerical results, such as the effective depletion factor of $^{137}$Cs specific activity.

The statistical procedure devised for this study, accounting also for values less than the MDA of the measuring operations, is of general use and can be used for other comparisons arising from radioactivity analyses.